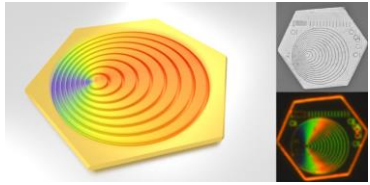

**Plasmonic Doppler Grating** provides azimuthal angle dependent continuous broadband lattice momentum and serves as a broadband continuous grating platform for micro- to nanophotonic spectroscopic applications.

# Design and Characterization of Plasmonic Doppler Grating for Azimuthal Angle-resolved Surface Plasmon Resonances

Kel-Meng See,[a][†] Fan-Cheng Lin,[a][†] and Jer-Shing Huang[a,b,c*]

a. Department of Chemistry, National Tsing Hua University, 101 Sec. 2, Kuang-Fu Road, Hsinchu 30013, Taiwan
b. Leibniz Institute of Photonic Technology, Albert-Einstein Str. 9, 07745 Jena, Germany.
c. Research Center for Applied Sciences, 128 Sec. 2, Academia Road, Nankang, Taipei 11529, Taiwan
† These authors contribute equally to this work.
* Email: jer-shing.huang@leibniz-ipht.de

## Abstract

We present two-dimensional plasmonic Doppler grating (PDG) for broadband azimuthal angle-resolved nanophotonic applications. The PDG consists of a set of non-concentric circular rings mimicking the wavefronts of a moving point source that exhibits Doppler Effect and thereby offers continuous azimuthal angle-dependent lattice momentum for photon-plasmon coupling. The center and span of the working frequency window is fully designable for optimal performance in specific applications. We detail the design, fabrication and optical characterization of the PDG. The design of Doppler grating provides a general platform for in-plane angle-resolved nanophotonic applications.

## Introduction

At the interface of dielectric and metal, electromagnetic energy can propagate in form of surface plasmons (SPs) with relatively large momentum compared to that of corresponding photons in vacuum[1]. To excite SPs, the momentum mismatch between the vacuum photons and the SPs must be compensated. Various schemes have been employed, including Kretschmann-Raether and Otto configurations with prisms[2,3], metallic gratings[4] and direct excitation of SPs via nonlinear wave-mixing on flat metal surfaces[5]. Among these excitation schemes, metallic gratings have been extensively used and investigated because of its simplicity in design and fabrication[4,6-9]. A grating can effectively promote the excitation of SPs because it contains periodically varying morphology that offers lattice momentum to fulfil the momentum matching condition[10,11],

$$\frac{2\pi}{\lambda_0} n_i \sin\alpha + \frac{2m\pi}{P} = \frac{2\pi}{\lambda_0}\sqrt{\frac{\varepsilon_m \cdot n_d^2}{\varepsilon_m + n_d^2}} \quad (1)$$

, where $\lambda_0$ is the wavelength of light in vacuum, $\alpha$ is the incident angle, $P$ is the periodicity, $m$ is the resonant order, $\varepsilon_m$ is the permittivity of the metal, $n_i$ is the refractive index of the medium, from which the light is impinging the grating, and $n_d$ denotes the index of the dielectric surrounding of the grating. Equation (1) links the resonant wavelength to the grating periodicity and provides a simple guideline for the design of grating SP couplers. For such simplicity, plasmonic gratings have been widely used for index sensing[6,12], color filtering[13-15], surface enhanced spectroscopy[16-18], hot electron generation[7], color selective CMOS photodetector[8], light tailoring and collimation[19] as well as logic gates in optical nanocircuits[20]. So far, most of the gratings are designed to have a constant periodicity and the performance is optimized at the resonant frequency. For multi-color and broadband applications, such as color sorting, color filtering and index sensing, SP couplers with multiple grating periodicities are needed. To this end, Laux et al.[13] have shown three-color sorting using hexagonal arrangement of triangular stacking groove arrays. The resonant frequencies are, however, discrete and the number of colors is limited to three. Xu et al.[14] proposed a 2D spoken structure as a nanosized plasmonic spectroscope for broadband continuous color sorting. With their structure, different colors are separated continuously in the radial direction. Therefore, the color sorting is not sensitive to the in-plane polarization of light. Furthermore, since the periodicity changes as a function of the radius, the spatial dispersion of color is automatically linearly proportional to the wavelength, rendering the design of spatial color dispersion impossible and the spectroscopic applications limited. In this work, we present a novel design of plasmonic Doppler grating (PDG), which provides continuous azimuthal angle-dependent periodicity for broadband plasmonic resonance at the optical frequencies. The center and span of the working frequency window can be freely designed to best fit the desired applications. Due to the azimuthal angle-dependent periodicity, PDG can function as an "index meter" that reports the variation of environmental index in the change of color distribution in azimuthal angle.

Here, we detail the theoretical design, experimental fabrication and optical characterization of the PDG and demonstrate an exemplary application in continuous broadband color sorting. PDGs may also find applications in index sensing, gas sensing, broadband plasmon-enhanced spectroscopy[16] and non-linear signal generation[21]. The design of Doppler grating is not limited to plasmonic materials but can also be applied to dielectric materials to fabricate, for example, broadband couplers or frequency routers in silicon photonic circuits or gratings for applications that require two-dimensional spatial dispersion of light[22-24].

## Experimental

### Design and Features of PDG

The PGD consists of a set of non-concentric circular rings describing the wave fronts of a moving point source that exhibits Doppler Effect. The trajectory of the $n^{th}$ ring can be mathematically expressed as $(x - nd)^2 + y^2 = (n\Delta r)^2$, where $\Delta r$ is the radius increment and $d$ is the ring centre displacement. The centre of each circular ring is displaced relative to the adjacent rings by constant displacement in one specific direction, mimicking the wavefronts from a point source moving with constant velocity. The distance between two adjacent rings is thus azimuthal angle-dependent and can be expressed as,

$$P(\varphi) = \pm d\cos\varphi + \sqrt{(d^2 \cos 2\varphi + 2\Delta r^2 - d^2)/2} \quad (2)$$

, where $\varphi$ is in-plane azimuthal angle. This distance defines the periodicity of the grating along certain azimuthal angle. The detailed derivation of the periodicity can be found in the *Supporting Information*. As shown in Fig. 1a, the whole PDG structure serves as a two dimensional grating that provides continuously varying periodicity as a function of in-plane azimuthal angle (φ). Combining Eq.2 with the momentum matching condition for surface plasmon excitation (Eq.1), we obtain the azimuthal angle-dependent surface plasmon coupling wavelength $\lambda_0$ as

$$\lambda_0 = \frac{\pm d\cos\varphi + \sqrt{(d^2 \cos 2\varphi + 2\Delta r^2 - d^2)/2}}{m}\left(\sqrt{\frac{\varepsilon_m \cdot n_d^2}{\varepsilon_m + n_d^2}} - n_i \sin\alpha\right). \quad (3)$$

For one specific single azimuthal angle, equation (3) yields two solutions of resonant wavelengths due to the two different periodicities. At $\varphi = 0°$, the two solutions correspond to the largest periodicity of PDG with a value of $\Delta r + d$ and the smallest one of $\Delta r - d$. As the azimuthal angle increases from 0° to 90°, the two periodicities gradually merge to one single value. The separation of the two spectral peaks at $\varphi = 0°$, therefore, defines the span of PDG's working spectral window, and the single spectral peak at $\varphi = 90°$ sets the spectral position of window. Since plasmonic gratings couples best to *p*-polarized light, the selection of azimuthal angle in real optical experiments can be easily done by using a polarizer in the excitation or detection beam path.

To achieve best performance of PDG, four design parameters can be freely tuned to optimize the position and the span of the working spectral window. These parameters are the radius increment, $\Delta r$, the center displacement, $d$, the material permittivity, $\varepsilon_m$, and the surrounding refractive index, $n_d$. Figures 1b-e illustrate the effect of tuning each of the parameters on the trend of the resonant wavelengths at various azimuthal angles. Clearly, the spectral position can be freely designed by choosing suitable values of $\Delta r$, $n_d$ and $\varepsilon_m$, and the span can be tuned by varying $d$. As shown in Fig. 1c, increasing the centre displacement $d$ results in an enlarged span of PDG's spectral window, suitable for applications concerning large spectral shift or covering wide spectral range. Decreasing the centre displacement, on the other hand, allows the PDG to "zoom in" to a small spectral range for ultimate sensitivity and resolution, as will be demonstrated in the exemplary application of PDG as an index meter. As for $n_d$ and $\varepsilon_m$, they can either be pre-designed by using different materials or be actively controlled by incorporating tunable materials whose optical properties can be externally tuned by light[25,26], heat[27] or electricity[28,29]. Alternatively, actively tunable PDG can also be achieved by using highly dosed semiconducting materials, whose permittivity can be controlled by external electric field[30-34]. It is worth noting that Fig. 1d indeed suggests that PDG can be used as an index meter, in which the variation of the environmental index can be translated into the change of in-plane color distribution with respect to the azimuthal angle, similar to the pointer in a speedometer. This makes PDG a simple and fully designable index meter for microfluidic lab-on-a-chip systems. In the following, we describe in detail the fabrication and optical characterization of PDG and demonstrate one application example of color sorting.

### Fabrication of PDG

To achieve the best performance in the visible to near-infrared spectral regime, we have chosen to fabricate PDGs on ultrasmooth chemically synthesized single-crystalline gold flakes using Gallium focused-ion beam milling. Briefly, the self-assembled gold flakes are synthesized using chemical method reported previously[35]. The solution of gold flakes is then drop-casted on a cover glass coated with a 40 nm thick ITO layer. The ITO glass contains pre-fabricated markers for target identification. The PDGs are fabricated using Gallium focused-ion beam milling (Helios Nanolab 600i System, FEI Company). The beam current and the acceleration voltage are 1.7 pA and 30 V, respectively. Single-crystalline gold flakes are chemically stable, atomically smooth and grain boundary free. These features make single-crystalline gold flake a perfect substrate for the fabrication of high-definition PDGs, which contain many fine features and extend over large area[35].

### Optical characterization of PDG

To characterize the optical response of the PDG, we have used home-built dark-field optical microscopes[36,37] to observe the images and spectra of PDGs in three different modes, namely the scattering mode, reflection mode and transmission mode. Results obtained from different observation modes provide us with complementary information to gain comprehensive insight into the optical response of the PDG. In the dark-field scattering experiment, white-light illumination (HAL 100 illuminator with quartz collector, Zeiss) is impinging the sample from the ITO substrate with Köhler illumination scheme using an oil condenser (Achromatic-aplanatic Condenser N.A. = 1.4, Zeiss). The scattered light is then collected by another objective (MPlanApo 60X air N.A. = 0.9, Olympus) on the side open to air and is aligned onto a color CCD (PMD-130, OME-TOP SYSTEMS CO., LTD.) to obtain color images and a spectrometer to record the azimuthal angle-resolved spectra. For the latter, two linear polarizers (LPVIS100-MP, 550-1500 nm, Thorlabs) are inserted into the excitation and detection optical pathways to select the azimuthal angle. Bandpass filters with finite bandwidth of 40 nm (FKB-VIS-40, Thorlabs) are used to obtain narrow band excitation centred at 550 nm, 600 nm, 650 nm and 750 nm. The optical characterization of PDG color sorter is performed both in reflection mode and in transmission mode. In the reflection mode, unpolarized white light source is used to illuminate the PDG with a nearly normal incident angle. The incident light and the reflected light are collected using the same air objective (DIN PLL 20 X, N.A. = 0.4, ZAK). For the transmission mode, the incident light is softly focused onto the sample with a microscope condenser (Air, N.A. = 0.3, Zeiss) and the transmitted light is collected by an oil objective (Plan-Apochromat 63X oil Iris, N.A. = 0.7-1.42, Zeiss). Schematic diagrams for the optical setups used in the three detection modes can be found in the Supporting Information.

**Numerical simulations on PDG**
Numerical simulations are performed with finite-difference time-domain method (FDTD Solutions, Lumerical Solutions). Dimensions used in the simulations are estimated according to the SEM images of the real PDG structures. The thickness of the ITO layer is about 40 nm. To simulate far-field scattering spectra for light scattering at different azimuthal angles, we perform two-dimensional simulations on the cross sectional planes perpendicular to the metallic surface at the azimuthal angle of interest. P-polarized total-field scattered-field plane wave source is injected from the glass layer with an incident angle of 55°, mimicking the experimental condition of the dark-field illumination. The scattering spectra were obtained by integrating the Poynting vector over a linear one-dimensional monitor placed in the air 1200 nm above the metallic surface. All simulated spectra have been normalized to the source spectrum to obtain the enhancement. To understand the resonance modes, the cross sectional magnetic near-field profile ($|H_z|$) of the PDG are recorded by a two-dimensional monitor that extends over air, gold and substrate layers. Total electric near-field intensity ($|E_x|^2 + |E_y|^2 + |E_z|^2$) profiles are obtained from three-dimensional FDTD simulations and are recorded on a two-dimensional plane that cuts through the middle height of the gold film. To simulate the optical response of a PDG under an unpolarized excitation, we have separately performed two simulations with orthogonal polarizations. The responses of the structures to the unpolarized excitation are then obtained by calculating the incoherent sum of the results from the two simulations according to the formula, $\langle |E|^2 \rangle = \frac{1}{2}|\vec{E}_s|^2 + \frac{1}{2}|\vec{E}_p|^2$.

## Results and discussion

**Optical response of PDG**
To cover the spectral range from ultraviolet to near infrared for optical frequency applications, a gold PDG with a radius increment $\Delta r$ = 400 nm and centre displacement $d$ = 200 nm has been fabricated. The fabricated gap width is about 45 nm and the thickness of the gold flake is about 70 nm. The schematic of the dark-field scattering optical experiment is depicted in Fig. 2a. The scanning electron microscope (SEM) image and the corresponding full color dark-field scattering image of the fabricated PDG are shown in Fig. 2b. As can be seen in Fig. 2b, the PDG shows clear azimuthal angle-dependent color dispersion in the visible spectral range. We have inserted band-pass color filters (bandwidth = 40 nm) in the excitation beam path to more clearly observe the intensity distributions as a function of azimuthal angle for 550 nm, 650 nm, and 750 nm. The corresponding intensity images are shown in the left panels in Fig. 2c. Upon illumination centred at 550 nm, two intensity maxima on the PDG are observed at $\varphi = \pm 83°$ and $\varphi = \pm 10°$. For illumination centred at 650 nm and 750 nm, the resonant modes are observed at $\varphi = \pm 48°$ and $\varphi = \pm 29°$, respectively. To understand the observed color distribution and their corresponding SP resonances, we have performed finite-difference time-domain numerical simulations to obtain the intensity distribution of the optical near field on a plane cutting the gold film at the middle height. Dimensions used in the simulations are estimated according to the SEM images.

The simulated near-field intensity distributions are displayed on the right panels in Fig. 2c. In general, the intensity angle distributions in the experimental scattering images are reproduced in the simulated near-field profile at all three wavelengths. However, some features seen in the simulated profiles are missing in the experimental far-field images. To understand the difference, it is important to keep in mind that the simulated distributions only reveal the near-field intensity profile inside the air slits. These modes seen in the near field do not necessarily couple to the far field unless the momentum matching condition is fulfilled. In addition, light with different wavelength is scattered to far field along different out-coupling angle. Therefore, the collection angle of the objective also matters. For out-coupling angle larger than the collection angle of the objective, the scattered light will not be detected by the CCD and spectrometer, resulting in a dark region in the scattering image. This is why the PDG appears dark at the region with small periodicity, as can be seen in the images in Figs. 2b and 2c. Details on the collection limitation of the objective are given in the *Supporting Information*.

The second reason for the difference between experimental and simulated results is about the excitation of SPs. Since the thickness of the gold film is only 70 nm, the illumination coming from the ITO half space can excite both SPs on the Au/ITO interface and SPs on the Au/Air interface. This is illustrated with the two waves in blue and red colors in Fig. 2a. The excited SPs on the two interfaces can then propagate to the other side via the vertical air slit and get scattered to the far field on the other interface. Since the excitation and detection are on the opposite sides of the gold layer (Fig. 2a), there are two possible routes for the scattered light to reach the detector. The first route is that the SPs are excited on the Au/ITO interface, propagate via the vertical slits to the other surface and get scattered to the detector by borrowing the momentum from the grating on the Au/Air interface. These SP modes are marked with white solid lines in Fig. 2c. In the second route, the SPs are directly excited on the Au/Air interface by the illumination coming from the ITO half space via Kretschmann-Raether excitation scheme[2]. Then, the SPs are scattered into the far field with the help of the lattice momentum provided by the gratings on the Au/Air interface. In this way, the SPs do not go into the vertical air slits. The excitation of SPs is much more efficient on the Au/ITO interface compared to that on the Au/Air interface because the illumination is coming from the ITO side. Also, the efficiency of the direct excitation of SPs on the Au/Air interface is very low also because the optical field is greatly attenuated by the metallic film. As a result, the first route contributes majorly to the intensity distribution in the experimental images and only the scattering signals from the SPs originally excited on the Au/ITO interface (first route) are observable in the experimental color images. Figure 2d shows the angle-resolved normalized spectra of the PDG obtained from dark-field scattering experiment (left panel) and FDTD simulations (right panel). Solid and dashed lines in different colors in Fig. 2d are obtained from an analytical model based on equation (3) and are duplicated on both panels for the convenience of comparison. These lines mark the analytically predicted spectral positions of various orders of SP modes excited on the Au/ITO interface (solid lines) and on the Au/Air interface (dashed lines). As can be seen, the peak positions in the experimental spectra show good agreement with the simulated spectra and the analytical prediction. Compared to the simulated spectra on the right panel of Fig. 2d, the peak shape is more broadened in the experimental spectra. This is due to the structural imperfection of the PDG structure. Another interesting feature that can be seen in Fig. 2d is the Fano line shape in both the experimental and simulated spectra. These Fano-like resonance peaks stem from the coupling of the in-plane SPs into the vertical air slits[6]. To understand the Fano resonances, we have simulated the optical magnetic field profiles |Hz| at the wavelengths of the Fano peaks (hollow triangles) and dips (solid triangles), as shown in the insets of Fig. 2d. These profiles provide direct information on the coupling mechanism and help us assign the modes. The coupling of the SP modes excited on the Au/Air interface (m = -1, blue dashed curves) into the vertical slit results in the spectral dip (blue solid triangle) because the SPs couple back to the opposite side without the detector. On the contrary, for the SP modes excited on the Au/ITO interface (via first route, m = -2, red dashed curves), the coupling of the SP modes into the vertical slit leads to the Fano peak (red solid triangle) because the SP modes can propagate via the vertical slit to the other interface and get scattered to the detector.

It is now clear that those modes seen in the simulated near-field distribution (right panel of Fig. 2c) but missing in the experimental color images (left panel of Fig. 2c) are the SP modes excited directly on the Au/Air interface but propagating back to the Au/ITO interface via the vertical slits. In some applications, such as index sensing, the direct excitation of SPs on the Au/Air interface is troublesome and needs to be minimized. This can be done by using gold flakes that are thick enough to sufficiently suppress Kretschmann-Raether excitation of SPs. This will greatly simplify the analysis of PDG's spectral response and facilitate the applications.

**Color sorting by PDG**

Having understood the optical response of the PDG, we now demonstrate the first exemplary application, a continuous in-plane polarization sensitive color sorter. For color sorting purpose, we have used a 400 nm thick gold flake and reduced the milling depth of FIB such that all rings except the middle one are milled to create grooves on the surface, instead of vertical slits that connects the two interfaces. For the middle ring, FIB has been applied to mill through the flake and create a vertical air slit (gap = ~100 nm) that connects the two interfaces. The main function of this middle ring is to serve as a sub-wavelength aperture that allows for transmission of SPs but rejects photons[10]. Figures 3a and 3b depict the cross sections of the PDG color sorter and the two detection geometries used for the optical characterization. The SEM image of the structure is shown in the inset of Fig. 3a. Such combination of a set of grooves and one vertical air slit has been shown to have superior color sorting efficiency[13]. More importantly, it allows us to observe the resonance of PDG color sorter in both reflection and transmission modes. Here, we have used bandpass filters (bandwidth = 40 nm) to create single band excitation centred at 550 nm, 600 nm and 650 nm. White light illumination without any filter has also been used to obtain the full-color images. Figures 3c and 3d show the reflection and transmission images on the left panels, respectively, with their corresponding intensity angle distribution profiles on the right panels. In the reflection images, dark bands are observed at various azimuthal angles depending on the wavelength. The dark bands in reflection images appear at the azimuthal angles, where the grating periodicity provides correct lattice momentum to fulfil the momentum matching condition required for the excitation of SPs. Therefore, the incident light is efficiently converted into near-field SPs and the intensity of the reflection is attenuated, resulting in a dark band. Since the sub-wavelength vertical air slit only allows the penetration of SPs but not photons[10], the SP resonances seen as dark bands in the reflection images manifest themselves as bright bands in transmission images. As can be seen in the bottom panel of Fig. 3d, the white light illumination is clearly sorted into a rainbow-like intensity distribution, confirming the azimuthal angle-dependent in-plane color sorting ability of the PDG. It is worth noting that the rainbow colors seen in the bottom panels of Figs. 3c (reflection) and 3d (transmission) are complementary. However, different from the color of plasmonic nanoparticle solutions, PDG's transmission directly reports the color of the SP resonances whereas the reflection presents the complementary color. The other feature is that the PDG color sorter is naturally sensitive to the polarization of incident light[13] since plasmonic gratings prefer *p*-polarized incident light. As shown in Fig. 4, the incident polarization controls the transmission color distribution. This provides additional degree of freedom to control the color sorting efficiency. It is worth noting that using a flake with proper thickness is important. The ideal flake thickness should be large enough to block the background due to the incident light, yet thin enough to allow the SPs to survive the loss of transmitting the slit. The transmission signals should be bright enough for observation. There are several possible applications of the PDG color sorter. For example, one can mill through only the centre ring (the smallest ring). In this case, the color rings shown in Fig. 4 shrink to points, which can function as pixels. By controlling the polarization of the white light illumination, we can control the relative intensity of the different color and thereby changing the overall color of the pixel. As a color sorter, the PDG can also serve as a designable on-site micro-spectrometer. For example, the PDG can sort different colors of the emitted light from a nanosized emitter into different azimuthal angle. Since photons with specific color can only couple to the free space through grating condition at specific azimuthal angles, one can obtain the emission spectrum by analysing the intensity angle distribution in the PDG image. This offers the possibility to perform spectroscopic analysis without using any farfield dispersive optical elements, such as prism or gratings, which is particularly useful for microfluidic or lab-on-a-chip analytical devices. The other possible application is to use Doppler grating as the coupler in silicon photonics as a multiplexer or a router. Upon broadband illumination, different frequency will be collected by the Doppler grating and coupled into different azimuthal angle. Multiple silicon waveguides arranged in different azimuthal angles can be fabricated under the Doppler grating to collect the coupled light. Photonic signals with different frequencies can also be merged through the Doppler grating.

**Quantitative analysis of azimuthal intensity distribution**

To precisely obtain the azimuthal angle of the SP resonance is very important for PDG's applications as analytical tools, e.g. PDG index meter or gas sensor, where the intensity angle distribution is quantitatively linked to the analytical information. For this purpose, we have performed quantitative analysis on the images in Fig. 3 and obtained the corresponding angle distribution profiles of the intensity, as shown on the right columns of Fig. 3c and 3d. To quantitatively determine the azimuthal angle of the resonance band, the reflection intensity angle distribution profiles are fitted with an analytical model based on Fano resonance model[38]. Details of the quantitative image analysis and fitting can be found in the *Supporting Information*. With the fitting, the grating periodicity for the $m^{th}$ order SP resonance, $P(\varphi_m)$, can be precisely determined. Consequently, by applying the fitted value of the periodicity to Eq. 2, we can quantitatively determine the azimuthal angle of the SP resonance from experiments. It is worth noting that the reflection or transmission angle distribution profiles of PDG have a finite bandwidth. This finite bandwidth stems from the broadening due to the damping of SPs, the degree of polarization of the illumination and the incident polarization angle with respect to the PDG grating edges. For sensing applications, it is very important to consider the broadening effect since the bandwidth limits the sensitivity. Therefore, we have included all the known broadening effects in the fitting model. By such analysis, we are able to obtain precise azimuthal angle of resonance and the uncertainty is mainly due to structural imperfection, which cannot be analytically modelled.

**Conclusions**

In conclusion, we have presented a design of PDG for continuous azimuthal angle-dependent plasmonic grating applications. By choosing correct $\Delta r$ and $d$, the centre and the span of the working spectral window of PDG can be easily designed to best fit the applications. We have characterized the angle resolved optical response by observing the scattering, reflection and transmission of various different PDG designs. Exemplary application of PDG as a continuous polarization sensitive broadband color sorter is demonstrated. The PDG can translate the change of environmental index into the change of intensity distribution in azimuthal angle and can find applications in index sensing. PDGs may also serve as a continuous grating structure for nonlinear signal generation or plasmon-enhanced spectroscopy. Using actively tunable materials, PDG's working frequency window can be actively controlled by external signals. For color sorting applications, the ability to active control promises the applications in actively tunable router or color pixel. The design of Doppler grating can also be applied to dielectric materials to create, for example, couplers for silicon photonic circuits or two-dimensional transparent gratings[22-24]. We anticipate applications of PDG in two-dimensional angle resolved spectroscopy and index sensing.

**Acknowledgements**

Supports from the Ministry of Science and Technology of Taiwan under Contract No. MOST-103-2113-M-007-004-MY3 is gratefully acknowledged. J.-S.H. thanks the support from Center for Nanotechnology, Materials Sciences, and Microsystems at National Tsing Hua University.

**References**

1. Barnes, W. L.; Dereux, A.; Ebbesen, T. W. Surface Plasmon Subwavelength Optics. *Nature* **2003**, *424*, 824-830
2. Kretschmann, E.; Raether, H. Radiative Decay of Non Radiative Surface Plasmon Excited by Light. *Z. Naturforsch. A* **1968**, *23a*, 2135-2136
3. Otto, A. Excitation of Nonradiative Surface Plasma Waves in Silver by the Method of Frustrated Total Reflection. *Z. Phys.* **1968**, *216*, 398-410
4. Schröter, U.; Heitmann, D. Surface-plasmon-enhanced Transmission through Metallic Gratings. *Phys. Rev. B* **1998**, *58*, 15419-15421
5. Renger, J.; Quidant, R.; van Hulst, N.; Palomba, S.; Novotny, L. Free-Space Excitation of Propagating Surface Plasmon

Polaritons by Nonlinear Four-Wave Mixing. *Phys. Rev. Lett.* **2009**, *103*, 266802

6. Lee, K. L.; Chen, P. W.; Wu, S. H.; Huang, J. B.; Yang, S. Y.; Wei, P. K. Enhancing Surface Plasmon Detection Using Template-Stripped Gold Nanoslit Arrays on Plastic Films. *ACS Nano* **2012**, *6*, 2931-2939

7. Sobhani, A.; Knight, M. K.; Wang, Y.; Zheng, B.; Kin, N. S.; Brown, L. V.; Fang, Z.; Nordlander, P.; Halas, J. N. Narrowband photodetection in the near-infrared with a plasmon-induced hot electron device. *Nat. Commun.* **2013**, *4*, 1643

8. Zheng, B. Y.; Wang, Y.; Nordlander, P.; Halas, N. J. Color-selective and CMOS-compatible Photodetection Based on Aluminum Plasmonics. *Adv. Mater.* **2014**, *26*, 6318-6323

9. López-Tejeira, F.; Rodrigo S. G.; Martín-Moreno, L.; García-Vidal, F. J.; Devaux, E.; Ebbesen, T. W.; Krenn, J. R.; Radko, I. P.; Bozhevolnyi, S. I.; González, M. U.; Weeber, J. C.; Dereux, A. Efficient Unidirectional Nanoslit Couplers for Surface Plasmons. *Nat. Phys.* **2007**, *3*, 324-328 2007

10. Ebbesen, T. W.; Lezec, H. J.; Ghaemi, H. F.; Thio T; A.Wolff, P. Extraordinary Optical Transmission through Sub-wavelength Hole Arrays. *Nature* **1998**, *391*, 667-669

11. Collin, S.; Pardo F.; Teissier R.; Pelouard, J. L. Strong Discontinuities in the Complex Photonic Band Structure of Transmission Metallic Gratings. *Phys. Rev. B* **2001**, *63*, 033107

12. Yeh, W. H.; Hillier, A. C. Use of Dispersion Imaging for Grating-coupled Surface Plasmon Resonance Sensing of Multilayer Langmuir-Blodgett Films. *Anal. Chem.* **2013**, 85, 4080-4086

13. Laux, E.; Genet, C.; Skauli, T.; Ebbesen, T. W. Plasmonic Photon Sorters for Spectral and Polarimetric Imaging. *Nat. Photon.* **2008**, *2*, 161-164

14. Xu, T.; Wu, Y. K.; Luo, X.; Guo, L. J. Plasmonic Nanoresonators for High-resolution Colour Filtering and Spectral Imaging. *Nat. Commun.* **2010**, *1*, 59

15. Zeng, B.; Gao, Y.; Bartoli, F. J. Ultrathin Nanostructured Metals for Highly transmissive Plasmonic Subtractive Color Filters. *Sci. Rep.* **2013**, *3*, 2840

16. Andrade, G. F. S.; Min, Q.; Gordon, R.; Brolo, A. G. Surface-Enhanced Resonance Raman Scattering on Gold Concentric Rings: Polarization Dependence and Intensity Fluctuations. *J. Phys. Chem. C* **2012**, *116*, 2672-2676

17. Lin, D.; Huang, J. S. Slant-gap Plasmonic Nanoantennas for Optical Chirality Engineering and Circular Dichroism Enhancement. *Opt. Express* **2014**, *22*, 7434-7445

18. Jiang, Y.; Wang, H. Y.; Wang, H.; Gao, B. R.; Hao, Y. W.; Jin, Y.; Chen, Q. D.; Sun, H. B. Surface Plasmon Enhanced Fluorescence of Dye Molecules on Metal Grating Films. *Phys. Chem. C* **2011**, *115*, 12636-12642

19. Aouani, H.; Mahboub, O.; Bonod, N.; Devaux, E.; Popov, E.; Rigneault, H.; Ebbesen, T. W.; Wenger, J. Bright Unidirectional Fluorescence Emission of Molecules in a Nanoaperture with Plasmonic Corrugations. *Nano Lett.* **2011**, *11*, 637-644

20. Fu, Y.; Hu, X,; Lu, C.; Yue, S.; Yang, H.; Gong, Q. All-optical Logic Gates Based on Nanoscale Plasmonic Slot Waveguides. *Nano Lett.* **2012**, *12*, 5784-5790

21. Wang, C. Y.; Chen, H. Y.; Sun, L.; Chen, W. L.; Chang, Y. M.; Ahn, H.; Li, X.; Gwo, S. Giant colloidal silver crystals for low-loss linear and nonlinear plasmonics. *Nat. Commun.* **2015**, *6*, 7734

22. Isailovic, D.; Li, H. W.; Phillips, G. J.; & Yeung, E. S. High-Throughput Single-Cell Fluorescence Spectroscopy. *Appl. Spectrosc.* **2005**, *59*, 221-226

23. Webb, M. R.; LaFratta, C. N.; Walt, D. R. Chromatically Resolved Optical Microscope (CROMoscope): A Grating-Based Instrument for Spectral Imaging. *Anal. Chem.* **2009**, *81*, 7309-7313

24. Goda, K.; Tsia, K. K.; Jalali, B. Serial time-encoded amplified imaging for real-time observation of fast dynamic phenomena. *Nature* **2009**, *458*, 1145-1149

25. Hsiao, V. K. S.; Zheng, Y. B.; Juluri, B. K.; Huang, T. J. Light-Driven Plasmonic Switches Based on Au Nanodisk Arrays and Photoresponsive Liquid Crystals. *Adv. Mater.* **2008**, *20*, 3528-3532

26. Dintinger, J.; Klein, S.; Ebbesen, T. W. Molecule–Surface Plasmon Interactions in Hole Arrays: Enhanced Absorption, Refractive Index Changes, and All-Optical Switching. *Adv. Mater.* **2006**, *18*, 1267-1270

27. Gehan, H.; Mangeney, C.; Aubard, J.; Lévi, G.; Hohenau, A.; Krenn, J. R.; Lacaze, E; Félidj, N. Design and Optical Properties of Active Polymer-Coated Plasmonic Nanostructures. *J. Phys. Chem. Lett.* **2011**, *2*, 926-931

28. Stockhausen, V.; Martin, P.; Ghilane, J.; Leroux, Y.; Randriamahazaka, H.; Grand, J.; Felidj, N.; Lacroix, J. C. Giant Plasmon Resonance Shift Using Poly(3,4-ethylenedioxythiophene) Electrochemical Switching. *J. Am. Chem. Soc.* **2010**, *132*, 10224-10226

29. Schaming, D.; Nguyen, V. Q.; Martin, P.; Lacroix, J. C. Tunable Plasmon Resonance of Gold Nanoparticles Functionalized by Electroactive Bisthienylbenzene Oligomers or Polythiophene. *J. Phys. Chem. C* **2014**, *118*, 25158-25166

30. Yi, F.; Shim, E.; Zhu, A. Y.; Zhu, H.; Reed, J. C.; Cubukcu, E. Voltage tuning of plasmonic absorbers by indium tin oxide. *Appl. Phys. Lett.* **2013**, *102*, 221102

31. Garcia, G.; Buonsanti, R.; Runnerstrom, E. L.; Mendelsberg, R. J.; Llordes, A.; Anders, A.; Richardson, T. J.; Milliron, D. J. Dynamically modulating the surface plasmon resonance of doped semiconductor nanocrystals. *Nano Lett.* **2011**, *11*, 4415-4420

32. Feigenbaum, E.; Diest, K.; Atwater, H. A. Unity-order index change in transparent conducting oxides at visible frequencies. *Nano Lett.* **2010**, *10*, 2111-2116

33. Brown, A. M.; Sheldon, M. T.; Atwater, H. A. Electrochemical Tuning of the Dielectric Function of Au Nanoparticles. *ACS Photonics* **2015**, *2*, 459-464

34. Naik, G. V.; Shalaev, V. M.; Boltasseva, A. Alternative plasmonic materials: beyond gold and silver. *Adv. Mater.* **2013**, *25*, 3264-3294

35. Huang, J. S.; Callegari, V.; Geisler, P.; Brüning, C.; Kern, J.; Prangsma, J. C.; Wu, X.; Feichtner, T.; Ziegler, J.; Weinmann, P.; Kamp, M.; Forchel, A.; Biagioni, P.; Sennhauser, U.; Hecht, B. Atomically flat single-crystalline gold nanostructures for plasmonic nanocircuitry. *Nat. Commun.* **2010**, *1*, 150

36. Chen W. L.; Lin, F. C.; Lee, Y. Y.; Li, F. C.; Chang, Y. M.; Huang, J. S. The Modulation Effect of Transverse, Antibonding, and Higher-Order Longitudinal Modes on the Two-Photon Photoluminescence of Gold Plasmonic Nanoantennas. *ACS Nano* **2014**, *8*, 9053-9062

37. Liu, H. W.; Lin, F. C.; Lin, S. W.; Wu, J. Y.; Chou, B. T.; Lai, K. J.; Lin, S. D.; Huang, J. S. Single-Crystalline Aluminum Nanostructures on a Semiconducting GaAs Substrate for Ultraviolet to Near-Infrared Plasmonics. *ACS Nano* **2015**, *9*, 3875-3886

38. Gallinet, B.; Martin, O. J. F. Influence of Electromagnetic Interactions on the Line Shape of Plasmonic Fano Resonances. *ACS Nano* **2011**, *5*, 8999-9008

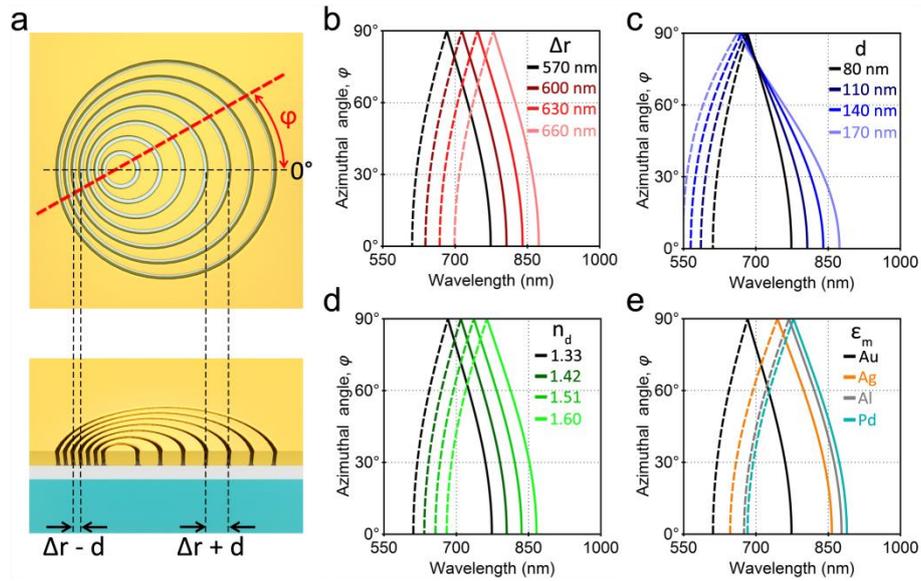

**Figure 1.** Design of the plasmonic Doppler grating. (a) Schematics illustrating the design of the PDG. Top panel shows the top view of a PDG with the azimuthal angle $\varphi$ marked in red dashed line. Lower panel shows the cross sectional view of a PDG on ITO glass. Extreme values of the periodicity at $\varphi = 0°$ are determined by the radius increment $\Delta r$ and the center displacement $d$. (b)-(e) Effects of varying four design parameters, (b) the radius increment $\Delta r$, (c) the center displacement $d$, (d) the refractive index of surrounding $n_d$ and (e) the permittivity of the PDG material. Solid and dashed lines mark the shifts of resonant wavelength on the long and short grating periods, respectively, as the azimuthal angle increases from 0° to 90°. Black lines in (b)-(e) are identical and are obtained analytically using the following parameter: $m = -2$, $d = 80$ nm, $\Delta r = 570$ nm, $n_d = 1.33$, $n_i = 1.33$, $\alpha = 48°$ and $\varepsilon_m$ for gold. These black lines serve as the reference for cross-figure comparison.

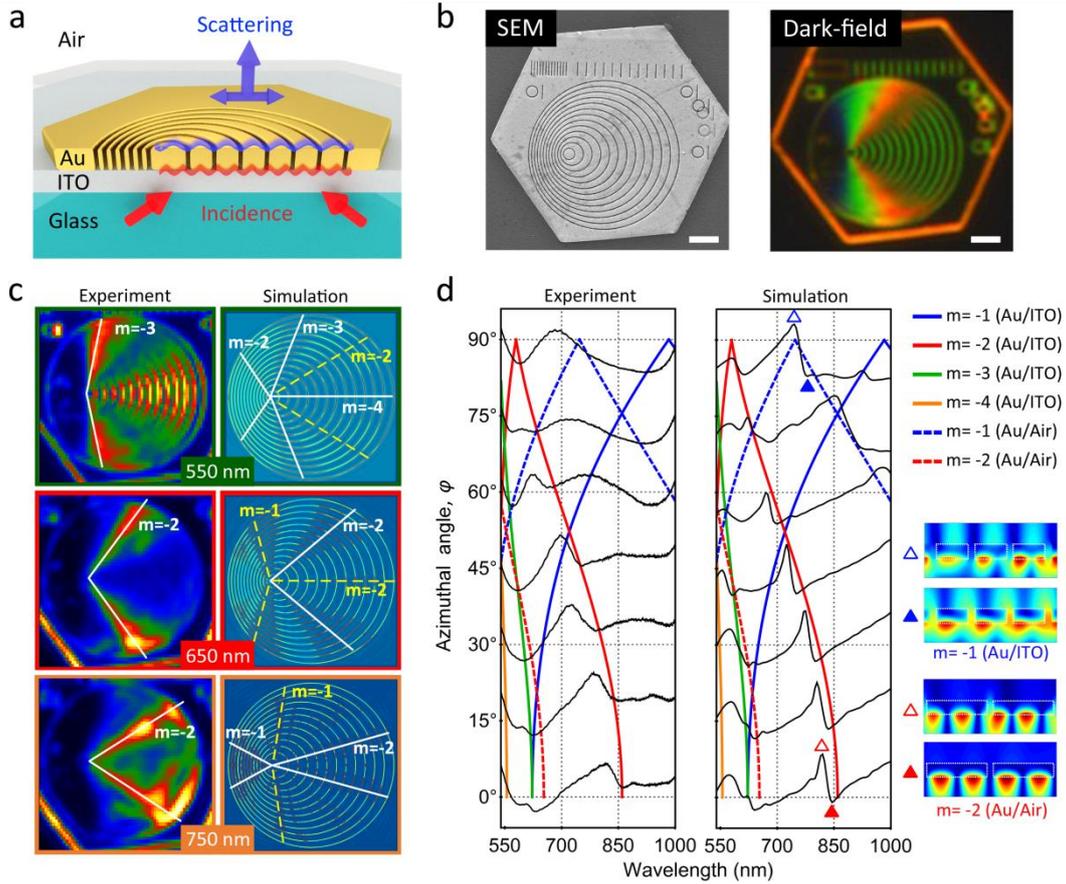

**Figure 2.** Optical characterization of PDG. (a) Schematic diagram illustrating the cross section of PDG and the configuration of dark-field microscope. Blue and red waves represent the SP modes excited on the two interfaces. (b) SEM image (left) and full color dark-field scattering image (right) of the PDG with $\Delta r$ = 400 nm and $d$ = 200 nm fabricated on a 70 nm thick single-crystalline gold flake. Scale bar: 2 µm. (c) Left panels: experimental dark-field scattering images recorded under narrow band excitation centered at 550 nm (top), 650 nm (middle), and 750 nm (bottom). Right panels: corresponding simulated profiles of electric field intensity ($E^2$) recorded on a plane cutting through the middle of the gold film. Intensity distributions from SP modes originally excited on the Au/ITO (solid lines) and Au/Air interface (dashed lines) are marked with the resonance order, m, as described in equation (1). (d) Azimuthal angle-resolved scattering spectra obtained from the dark-field scattering experiment (left panel) and the FDTD simulations (right panel). Analytical prediction of the azimuthal angle-dependent SP resonances on the Au/ITO (solid lines) interface and Au/Air interface (dashed lines) are marked with lines in blue, red, green and yellow color for m = -1, -2, -3, -4, respectively, as indicated in the legends. The insets show the simulated magnetic near-field profile (|Hz|) of the m = -1 mode on the Au/Air interface and the m = -2 mode on the Au/ITO interface. Hollow and solid triangles mark the wavelengths, at which the magnetic near-field profiles (|Hz|) in the insets are recorded.

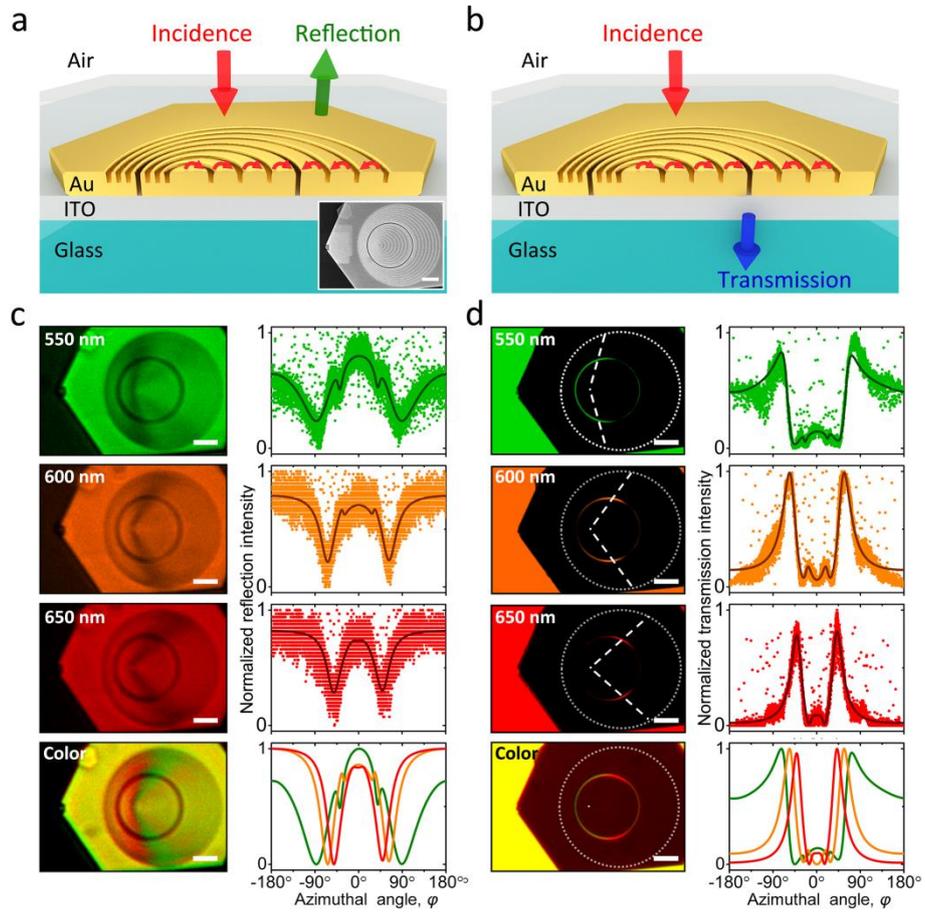

**Figure 3.** Application of PDG as a color sorter in reflection and transmission mode. (a) and (b) show the schematic diagrams illustrating the optical characterization in reflection mode and transmission mode, respectively. Inset: SEM image of the PDG color sorter with design parameters $\Delta r$ = 400 nm and $d$ = 200 nm. (c) From top to bottom shows the reflection color images (left panels) and the reflection intensity profile (right panels) taken under 550 nm, 600 nm, 650 nm and white light illumination. The solid lines are obtained by fitting the experimental data with an analytical function based on Fano resonance model. (d) Same as the images and plots in (c) but in transmission mode. All scale bars are 3 μm.

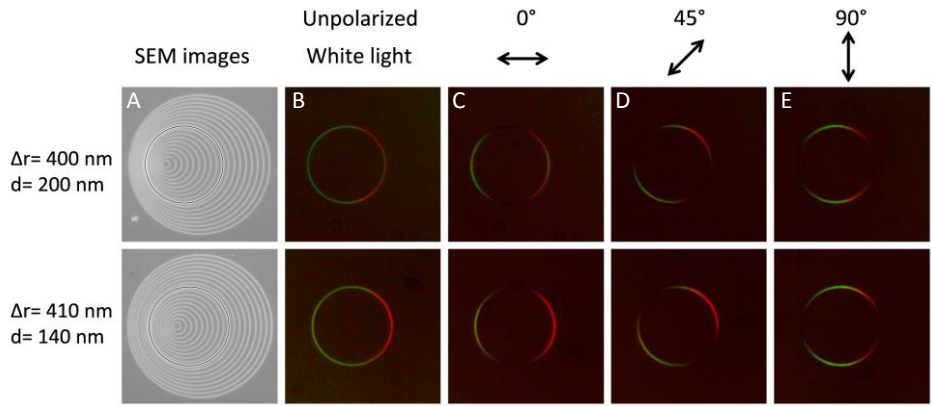

**Figure 4**. Polarization dependent color sorting. (A) SEM images of the PDG with different design parameters. (B)-(E) Transmission images of the rainbow color ring of the PDG color sorters under (B) unpolarized illumination and (C)-(E) linearly polarized illumination with polarization angle of 0°, 45° and 90°, respectively.